\documentclass[%
 prl,
 amsmath,amssymb,
 reprint,%
]{revtex4-1}

\usepackage{graphicx}%
\usepackage{dcolumn}%
\usepackage{bm}%
\usepackage{xcolor}
\usepackage{siunitx}

\newcommand{\tr}{\mathrm{tr}}

\newcommand{\ii}{\mathrm{i}}

\newcommand{\thalf}{\ensuremath{\tfrac{1}{2}}}
\newcommand{\bs}{\ensuremath{\bar{\bm{s}}}}
\usepackage{soul} %

\usepackage{braket}

\usepackage{dcolumn} %
\newcolumntype{d}[1]{D{.}{.}{#1}} %

\begin{document}

\title{%
Quantum entanglement from classical trajectories}

\author{Johan E. Runeson}
\email{johan.runeson@phys.chem.ethz.ch}
\author{Jeremy O. Richardson}%
\email{jeremy.richardson@phys.chem.ethz.ch}
\affiliation{Laboratory of Physical Chemistry, ETH Z\"{u}rich, 8093 Z\"{u}rich, Switzerland}

\date{\today}%

\begin{abstract}
A long-standing challenge
in mixed quantum-classical trajectory simulations is the treatment of entanglement between the classical and quantal degrees of freedom. We present a novel approach which describes the emergence of entangled states entirely in terms of independent and deterministic Ehrenfest-like classical trajectories.
For a two-level quantum system in a classical environment, this is derived
by mapping the quantum system onto a path-integral representation of a spin-$\thalf$. %
We demonstrate that the method correctly accounts for coherence and decoherence and thus reproduces the splitting of a wavepacket in a nonadiabatic scattering problem. This discovery opens up a new class of simulations as an alternative to stochastic surface-hopping, coupled-trajectory or semiclassical approaches.
\end{abstract}

\maketitle

\emph{Introduction}.---Many important phenomena across physics and chemistry are best described by a small quantum system and a large classical environment, for example light-matter interaction, chemical reactions, and qubits.
As it is intractable to treat the entire problem with quantum mechanics, it is necessary to simulate the coupled quantum-classical dynamics directly \cite{Stock2005nonadiabatic}.
Deriving an approach which is both computationally efficient and accurate is, however, a highly non-trivial task.
The simplest method based on classical trajectories is Ehrenfest dynamics, also known as mean-field theory (MFT).
While this approach is computationally efficient, it completely neglects quantum-classical entanglement, %
such as the branching of a nuclear wavepacket in a nonadiabatic scattering problem \cite{Tully2012perspective}.

Over the last decades, a considerable effort has been invested in the development of more accurate trajectory-based methods. A popular approach, especially in simulations of photochemistry, is Tully's fewest switches surface hopping (FSSH) \cite{Tully1990hopping,Subotnik2016review,prezhdo2005}, whose trajectories take stochastic jumps to simulate wavepacket branching. Although its original form is known to suffer from overcoherence, there have been many suggestions to introduce decoherence corrections \cite{subotnik2011AFSSH, prezhdo2016perspective, *granucci2007} with little consensus that there is a definitive solution.
Another known way to include entanglement is to use coupled trajectories, either on top of Ehrenfest \cite{shalashilin2011} or surface hopping \cite{martens2019} or through methodologies such as the exact-factorization framework \cite{abedi2010EF,Min2015nonadiabatic,agostini2016coupled}, ab-initio multiple spawning \cite{Curchod2018review}, the quantum-classical Liouville equation \cite{Kelly2012mapping},
or Bohmian dynamics \cite{prezhdo2001bohmian,*Curchod2013NABDY}.
A third possibility is to use interference between path histories and weight Ehrenfest-like trajectories (obtained from a mapping scheme \cite{mccurdy1978, Meyer1979nonadiabatic, Stock1997mapping} which has a close relation to the Stratonovich--Weyl representation used in the present paper \cite{runeson2019,runeson2020}) by phases and prefactors derived from a semiclassical propagator based on a real-time path integral \cite{Sun1997mapping,Miller2009mapping, miller2012perspective,makri2011forward}.
At first sight, decoherence and entanglement appear to be inherently quantum phenomena which cannot be described with a fully classical simulation \cite{ollitrault2020prl}.
However, in this Letter, we introduce a new approach that, in contrast to the three approaches described above, can capture these effects based on independent and deterministic classical trajectories.

Our theory is based on the Stratonovich--Weyl (SW) phase-space representation of the quantum system, which is a Wigner representation of discrete spaces \cite{brif1999phase,klimov2009group}. %
For simplicity, we consider only
the two-level case, which employs the well-known isomorphism to a spin $S=\thalf$ system, representing the spin by a classical vector of length $\sqrt{S(S+1)}$. We propose to extend this approach to a path integral of spin vectors, where the centroid of the spin path determines the dynamics and the initial configuration specifies the weight of each trajectory. This weight, which is preserved along the trajectory, contains the information necessary for quantum-classical entanglement. %

\emph{Method}.---%
First, consider an isolated two-level quantum system with density matrix $\hat{\rho}$.
A convenient classical analogue for this system is given by the Stratonovich--Weyl W-representation \cite{stratonovich1957distributions}, which expresses the expectation value of an operator $\hat{A}$ as an integral,
\begin{equation} \label{eq:rhoA}
    \tr[\hat{\rho}\hat{A}] = \int d^2\bm{s} \, \rho(\bm{s})A(\bm{s}).
\end{equation}
The classical functions are defined as $\rho(\bm{s})=\tr[\hat{\rho}\hat{w}(\bm{s})]$ (and likewise for $A(\bm{s})$) where $\hat{w}(\bm{s})=\frac{1}{2}\hat{\mathcal{I}} + \bm{s}\cdot\hat{\bm{\sigma}}$ is the SW kernel, $\hat{\mathcal{I}}$ is the $2\times 2$ identity matrix, $\hat{\bm{\sigma}}=[\hat{\sigma}_x,\hat{\sigma}_y,\hat{\sigma}_z]$ are the Pauli matrices, and $\bm{s}$ is a vector with magnitude $|\bm{s}|=\frac{\sqrt{3}}{2}$. For the integration measure we use the shorthand notation $\int d^2\bm{s}=\frac{1}{2\pi}\int d\varphi\,d\theta\, \sin\theta$, where $\varphi$ and $\theta$ are the spherical coordinates of $\bm{s}$. Since each Cartesian component, $s_j$, is the SW representation of the spin operator $\hat{S}_j=\thalf\hat{\sigma}_j$, one can think of $\bm{s}$ as a classical spin vector with the familiar quantum magnitude $\sqrt{S(S+1)}$ of a spin $S=\frac{1}{2}$ (where $\hbar=1$ throughout).

Next, consider the time evolution of the density matrix. As is well known, the dynamics of a two-level system is equivalent to that of a spin-$\thalf$ in an effective magnetic field $\bm{H}$, where the Hamiltonian is $\hat{H} = H_0\hat{\mathcal{I}}+ \bm{H}\cdot\hat{\bm{S}}$.
Using this decomposition, it is straightforward to write $H(\bm{s}) = H_0 + \bm{H}\cdot\bm{s}$ and likewise $\rho(\bm{s})=\rho_0 + \bm{\rho}\cdot\bm{s}$, where $\rho_0=\thalf$ is fixed by the normalization. When the Liouville--von-Neumann equation, $\frac{d}{dt}{\hat{\rho}} = \ii [\hat{\rho},\hat{H}]$, is converted to its phase-space equivalent,
\begin{equation}\label{eq:vonNeumann}
    \frac{d}{d t}\rho(\bm{s}) %
    = \ii\; \tr[(\hat{\rho}\hat{H}-\hat{H}\hat{\rho}) \hat{w}(\bm{s})] = \bm{\rho}\cdot(\bm{s}\times\bm{H}),
\end{equation}
it follows that the standard precession formula for the classical spin vector, $\dot{\bm{s}}=\bm{s}\times\bm{H}$, generates the correct quantum dynamics. %

When coupled to a general classical environment (described by coordinates $x$, mass $m$ and conjugate momenta $p$), the total Hamiltonian,
\begin{equation}
    \hat{H} = \left(\frac{p^2}{2m} + U(x)\right)\hat{\mathcal{I}} + \begin{pmatrix} V_1(x) & \Delta^*(x) \\ \Delta(x) & V_2(x) \end{pmatrix},
\end{equation}
corresponds to $H_0(x,p)=\tfrac{p^2}{2m}+U(x) + \frac{1}{2}[V_1(x)+V_2(x)]$ and $\bm{H}(x)=[2\,\text{Re}\,\Delta(x),2\,\text{Im}\,\Delta(x),V_1(x)-V_2(x)]$. %
The corresponding equations of motion are \cite{runeson2019}
\begin{equation}\label{eq:eom_xp}
    \dot{x} = \frac{p}{m}, \qquad \dot{p}= - \frac{\partial H_0}{\partial x} - \frac{\partial \bm{H}}{\partial x}\cdot \bm{s}
\end{equation}
in addition to the spin dynamics as before. While these equations of motion are equivalent to those of Ehrenfest dynamics \cite{meyer1979spin}, the SW treatment differs in the initial distribution: while standard Ehrenfest starts from a unique vector $\bm{s}$ of length $\thalf$ (as in the Bloch-sphere picture), the SW approach averages over all initial spin directions in Eq.~\eqref{eq:rhoA} and uses the magnitude $\frac{\sqrt{3}}{2}$. We have recently found that the latter, called the \emph{linearized spin-mapping method}, leads to a better prediction of population dynamics %
\cite{runeson2019,runeson2020,spinPLDM1,*spinPLDM2}. Other mapping approaches have also found
an effective spin magnitude of $\frac{\sqrt{3}}{2}$ to be optimal \cite{Mueller1998mapping,Cotton2013mapping}, and averaging over initial directions to be beneficial  \cite{fay2019faraday},
even with the Ehrenfest spin length \cite{shalashilin2019CSS,*liu2019}.
However, one important drawback is present in both Ehrenfest and linearized spin mapping, namely that the dynamical quantization is lost. %
This has the unfortunate consequence that %
after a scattering event,
the trajectories evolve on a weighted average of the two product potential energy surfaces, in contrast with the correct entangled state which splits into parts on one or the other surface \cite{Tully2012perspective}.
We will now show that such quantization can be systematically reintroduced by representing the system by a path integral of spins. %
In contrast to standard spin coherent-state path integrals, we do not require paths to be continuous in the $N\to\infty$ limit and therefore do not have to deal with the difficulties that arise when restricting to such paths \cite{SchulmanBook,altland2010condensed,wilson2011breakdown}.

By construction, the SW representation has an inversion formula
\begin{equation} \label{eq:rhoint}
    \hat{\rho} = \int d^2\bm{s}\, \rho(\bm{s}) \hat{w}(\bm{s}),
\end{equation}
with the particular example of the identity,
$\hat{\mathcal{I}} = \int d^2\bm{s}\, \hat{w}(\bm{s})$. By applying Eq.~\eqref{eq:rhoint} to both operators in $\tr[\hat{\rho}\hat{A}]$ %
and inserting resolutions of the identity, we can generalize Eq.~\eqref{eq:rhoA} to a path integral of $N$ spins,
\begin{multline}\label{eq:rhoA2}
    \tr[\hat{\rho}\hat{A}] = \int \left(\textstyle\prod_{k=1}^N d^2\bm{s}_k\right) \tr\left[\textstyle\prod_{k=1}^N \hat{w}(\bm{s}_k)\right] \times \\
    \textstyle \frac{1}{N}\sum_{l=1}^N \rho(\bm{s}_l)\,\frac{1}{N}\sum_{m=1}^N A(\bm{s}_{m}),
\end{multline}
where we symmetrized over the indices $l$ and $m$
and used Eq.~\eqref{eq:rhoA} for terms with $l=m$.
Due to the linearity of the SW representation, it follows that $\frac{1}{N}\sum_l \rho(\bm{s}_l) = \rho(\bar{\bm{s}})$ (and similar for $A$), where we introduced the centroid $\bar{\bm{s}}=\frac{1}{N}\sum_l \bm{s}_l$.
The expression looks like a classical phase-space average with a weight function $g_N(\{\bm{s}_k\}) \equiv \tr\left[\prod_k \hat{w}(\bm{s}_k)\right]$.
Note that if we had used $|\bm{s}|=\frac{1}{2}$, the weight function would reduce to that of standard spin coherent-state path integrals %
\cite{Kleinert}. However, we find that our choice $|\bm{s}|=\frac{\sqrt{3}}{2}$ converges quicker in $N$. %

A practical consideration is that $g_N(\{\bm{s}_k\})$ is a complicated complex-valued function that varies rapidly for high $N$. %
However, since the observables depend only on $\bar{\bm{s}}$ and not on the relative geometry, it is possible to rigorously integrate all degrees of freedom other than the centroid. Explicitly, we define
\begin{equation}
     G_N(\bs) \equiv \int \left(\textstyle\prod_{k=1}^N d^2\bm{s}'_k\right) g_N(\{\bm{s}'_k\}) \delta^{(3)}(\bs-\bar{\bm{s}}').
\end{equation}
Note that the weight function $g_N(\{\bm{s}_k\})$ is invariant under global rotations of the spin vectors \footnote{Let $\hat{U}$ be a unitary representation of the rotation $\bm{s}\to \bm{s}'$, then $\hat{w}(\bm{s}')=\hat{U}\hat{w}(\bm{s})\hat{U}^{-1}$ and the invariance follows from the cyclicity of the trace.}, so that $G_N$ is spherically symmetric, $G_N(\bs)=G_N(\bar{s})$, where $\bar{s}=|\bs|$.
Equation~\eqref{eq:rhoA2} thus simplifies to
\begin{equation}\label{eq:rhoA3}
    \tr[\hat{\rho} \hat{A}] = \int d^3\bar{\bm{s}} \, G_N(\bar{s}) \rho(\bar{\bm{s}}) A(\bar{\bm{s}}).
\end{equation}
Since the centroid of $N>1$ points on a sphere can reach any point inside the sphere, the integration domain of $\bs$ is the ball $|\bar{\bm{s}}|\leq \frac{\sqrt{3}}{2}$. For $N=1$, we define $G_1(\bar{s})=\frac{2}{3\pi}\delta(\bar{s}-\sqrt{3}/2)$, which recovers Eq.~\eqref{eq:rhoA}. %

\begin{figure}
    \centering
    \includegraphics{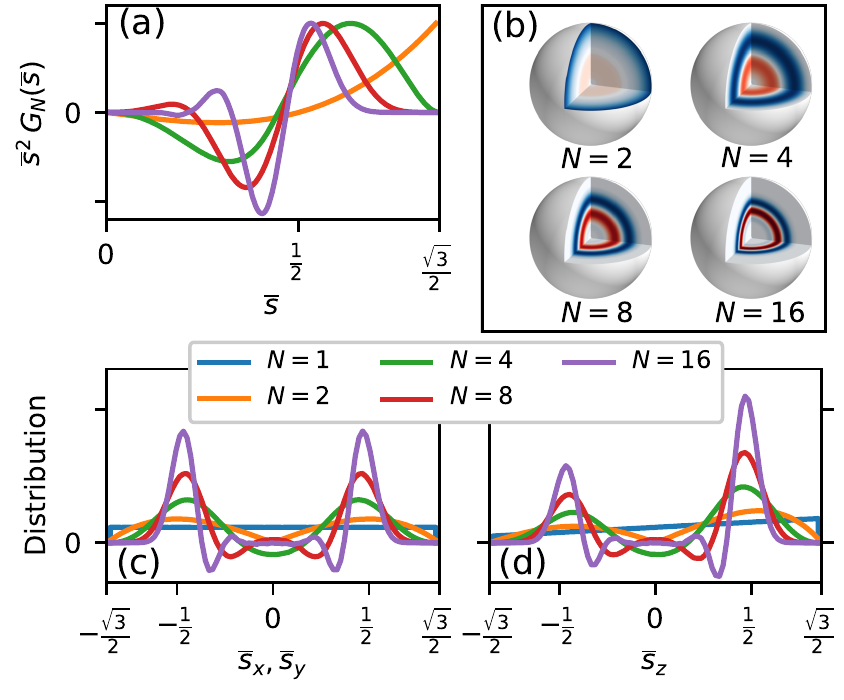}
    \caption{(a) Weight of the centroid spin magnitude $\bar{s}$ (arbitrary scaling). The negative regions grow in importance for increasing number of spins, $N$, but the number of nodes remains small. %
    (b) Weight of $\bar{\bm{s}}$ with positive contributions in blue and negative in red. (c-d) Distribution of the $x$, $y$ (c) and $z$ (d) components of the spin centroid. The distributions become peaked around the quantum-mechanical values $\pm \thalf$ with increasing $N$, and the relative peak heights approach the corresponding expectation values of the density matrix, here plotted for $\hat{\rho}=\frac{2}{3}|1\rangle\langle 1|+\frac{1}{3}|2\rangle\langle 2|$.}
    \label{fig:univ}
\end{figure}

The resulting \emph{universal function} $G_N(\bar{s})$ has several important properties: %
(1) it depends only on the centroid magnitude not on its direction, (2) it is real-valued, (3) it is independent of the Hamiltonian and the initial conditions. In other words, even though its computation becomes exponentially hard with increasing $N$, it only has to be computed once for a given $N$, hence the name universal. We have evaluated $G_N(\bar{s})$ numerically up to $N=16$ using Monte Carlo \cite{suppl_mat}. %
Figure~\ref{fig:univ}(a-b) shows that the universal function consists of a few positive and negative domains, but the number of nodes seems to remain small for high $N$. The simulation will thus include trajectories with both positive and negative weights but this does not lead to a severe sign problem  \cite{suppl_mat}.

Next, we consider the distribution of the spin components, $\hat{S}_j$. %
Quantum-mechanically these are expected to be quantized with the eigenvalues $\pm\thalf$, but the integrand in Eq.~\eqref{eq:rhoA} is smeared over all spin directions. %
However, as $N$ increases, the centroid distribution of Eq.~\eqref{eq:rhoA3} becomes peaked around $\bar{s}_j=\pm\thalf$ for all $j\in\{x,y,z\}$, with heights that are consistent with the components of $\hat{\rho}$, as shown in Figure~\ref{fig:univ}(c-d). In other words, the path-integral weight function $G_N(\bar{s})$ reintroduces the quantization to the system that is necessary for quantum-classical entanglement.

Finally, consider time-dependent expectation values. Using a similar argument as in Eq.~\eqref{eq:vonNeumann}, one can describe the dynamics in the spin path-integral representation by a homogeneous precession of all spins, $\dot{\bm{s}}_k=\bm{s}_k\times \bm{H}$. Consequently, the centroid evolves in the same way, $\bar{\bm{s}}=\bar{\bm{s}}\times\bm{H}$ and we do not need to keep track of the individual spin vectors. Since $G_N(\bar{s})$ is invariant under global rotations, its value is \emph{preserved} by the dynamics, which has the important implication that Eq.~\eqref{eq:rhoA3} is valid for all times. %

For an isolated system, this %
gives exact time-dependent expectation values for any value of $N$.
For the coupled quantum-classical problem, we propose the approximation
\begin{multline}\label{eq:rhotA}
    \tr[\hat{\rho} \hat{A}(t)]
    \approx \int dx \, dp \, d^3\bar{\bm{s}} \, G_N(\bar{s}) \rho(x,p,\bar{\bm{s}}) A(x_t,p_t,\bar{\bm{s}}_t),
\end{multline}
where the phase-space version of the density operator $\rho(x,p,\bar{\bm{s}})$ involves a Wigner transform of the environment in addition to the SW transform of the quantum system (and likewise for $A$).
This equation is the main result in this Letter and will be referred to as the \emph{spin path-integral method}.
It is exact at $t=0$ and in the limit of an isolated system for all $N$.
The $N=1$ case uses the same dynamics and spin distribution as the linearized spin-mapping method \cite{runeson2019} and in this more general formula, the accuracy is expected to increase with $N$ due to the quantization of the spin vectors.

\emph{Results}.---We have applied the spin path-integral method to Tully's seminal scattering problems \cite{Tully1990hopping}, which are well-known benchmark models but also proxies for realistic chemical reactions \cite{ibele2020tully}. The results are compared against calculations using numerically exact quantum mechanics as well as Ehrenfest dynamics and surface hopping. %
Simulation details are included in the Supplemental Material (SM) \cite{suppl_mat}.

\begin{figure}
    \centering
    \includegraphics{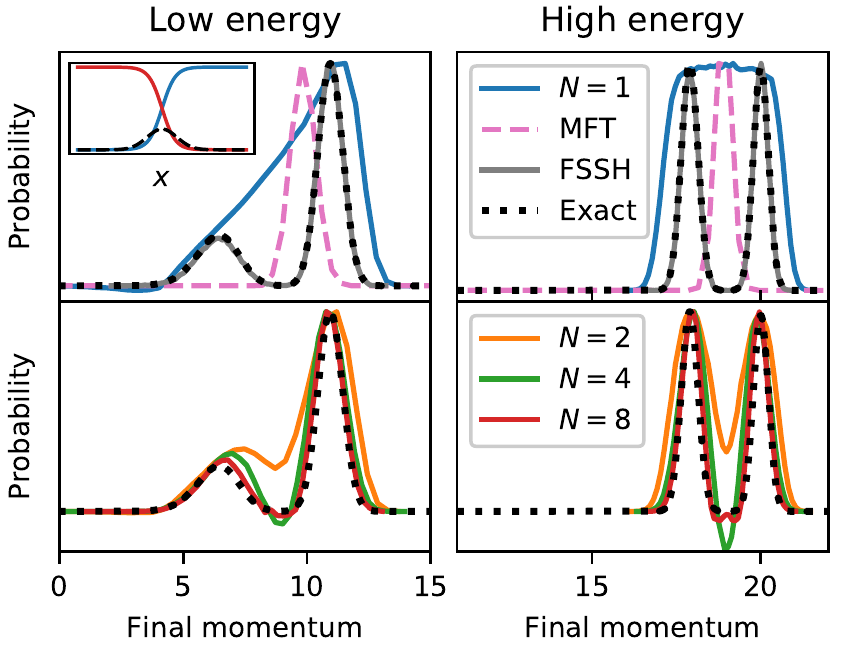}
    \caption{Probability distribution of the nuclear momentum after an avoided crossing (model I).  The inset shows the diabatic potentials (solid lines) and coupling (dashed line). A wavepacket enters from the left on the lower surface with a narrow distribution of kinetic energies at roughly 1.5 (left panels) or 5 (right panels) times the asymptotic energy difference. Ehrenfest (MFT) gives a single peak around the average momentum, while linearized spin mapping ($N=1$) envelopes the exact wavepacket distribution. For higher $N$, the spin path-integral method correctly reproduces the wavepacket branching.}
    \label{fig:symcross}
\end{figure}

First, we consider the single avoided crossing (model I) with diabatic surfaces and coupling shown in the inset of Fig.~\ref{fig:symcross}. %
An initial Gaussian wavepacket enters from the left on the lower surface, \mbox{$\psi_1(x,t=0)\otimes|1\rangle$}, with enough kinetic energy that both product channels are open. Due to nonadiabatic coupling, the wavepacket splits into two separate wavepackets on the two surfaces and emerges as an entangled state, ${\psi_1(x,t=\infty)\otimes|1\rangle} + {\psi_2(x,t=\infty)\otimes|2\rangle}$, on the right.
By \emph{quantum-classical entanglement}, we mean that if the nuclei are in a certain region of phase space, then we know with certainty the quantum state of the electrons and vice versa.
In Fig.~\ref{fig:symcross} we show the distribution of the final momentum for two different initial energies. %
As is well known, Ehrenfest dynamics (MFT) is unable to capture the branching of the wavepacket, whereas surface hopping (FSSH) provides a reasonably accurate description for this model. The $N=1$ simulation predicts an envelope that covers the full range of momenta allowed by energy conservation, but lacks the two-peak structure. However, by increasing $N$, we find that the distribution smoothly splits into two parts and thus recovers the quantum-mechanical entanglement.%

We emphasize that the dynamics consists of independent and deterministic trajectories on a weighted average of the two states, similar to both Ehrenfest dynamics and the linearized %
spin-mapping method, and the key difference lies in the weighting of the trajectories. Because the weights may be positive or negative, some of these cancel out in such a way that the ensemble branches when it emerges on uncoupled surfaces. This cancellation is reminiscent of more involved semiclassical methods such as Miller's forward-backward propagator, which is also known to capture wavepacket splitting in the present model \cite{Sun1997mapping,Miller2009mapping, miller2012perspective}.
However, these approaches are inherently semiclassical, not classical, and include nuclear-coherence effects (to some level of approximation)
via phases and prefactors that %
depend sensitively on the trajectory histories and make sampling difficult.
The results of the simpler spin path-integral method
demonstrate that only electronic coherence is necessary to recover the correct result.
Although the trajectories also carry a sign, this depends in a relatively simple manner on a single degree of freedom, is fixed by the initial sampling, and is preserved by the dynamics.

For Tully's dual avoided crossing (model II) we reach the same conclusions, and in the SM we show that the scattering probabilities are in good agreement with exact wavepacket calculations for a wide range of initial momenta \cite{suppl_mat}.

\begin{figure}
    \centering
    \includegraphics{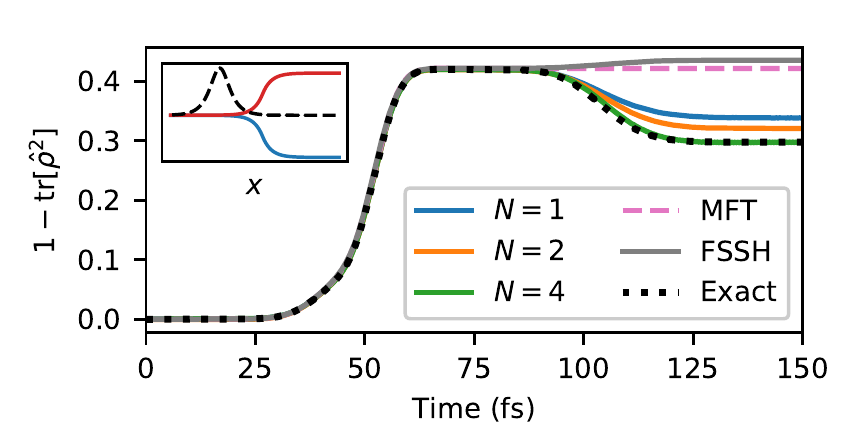}
    \caption{Impurity in the extended coupling model (model III). The inset shows the adiabatic surfaces (solid) and nonadiabatic coupling (dashed). A wavepacket enters from the left and is partly reflected.
    Only the spin path-integral method with $N=4$ is able to correctly describe the second crossing of the interaction region (results for $N=8$, not shown, overlay with $N=4$).}
    \label{fig:tully3_imp}
\end{figure}

Next, consider the more challenging extended coupling model (model III) shown in the inset of Fig.~\ref{fig:tully3_imp}.
As before, an initial wavepacket enters on the lower surface from the left but now the total energy is low enough for the upper channel to be closed on the right. During the collision, it thus splits into a transmitted part on the lower surface and a part on the upper surface which reflects and passes through the interaction region a second time. Surface hopping is well known to fail dramatically for systems with recrossing, because the electronic amplitudes picked up during the first crossing are inconsistent with the active surfaces \cite{Subotnik2016review}. %
This `overcoherence' problem arises because the assumption of a unique trajectory for each electronic density matrix is not valid \cite{subotnik2013QCLE} and is related to neglecting quantum-classical entanglement.  %
Ehrenfest and various linearized mapping approaches have also been unable to describe this model \cite{linearized}.

To quantify the overcoherence problems of quantum-classical simulations, we have calculated the time evolution of the impurity $S_L=1-\tr[\hat{\rho}^2]=2(\rho_{11} \rho_{22}-|\rho_{12}|^2)$, which is a measure of entanglement
and is related to the decoherence indicator studied in Ref.~\onlinecite{Min2015nonadiabatic} with coupled-trajectory simulations.  Here, $\rho_{nm}$ denotes elements of the reduced density matrix in the \emph{adiabatic} representation and the results are shown in Fig.~\ref{fig:tully3_imp}. Ehrenfest completely misses the second crossing at about 100 fs (since its trajectories do not reflect), and although some surface-hopping trajectories do reflect, FSSH is unable to correctly describe the entanglement in this system. The spin path-integral method on the other hand reproduces the correct result for this system.

\begin{figure}
    \centering
    \includegraphics{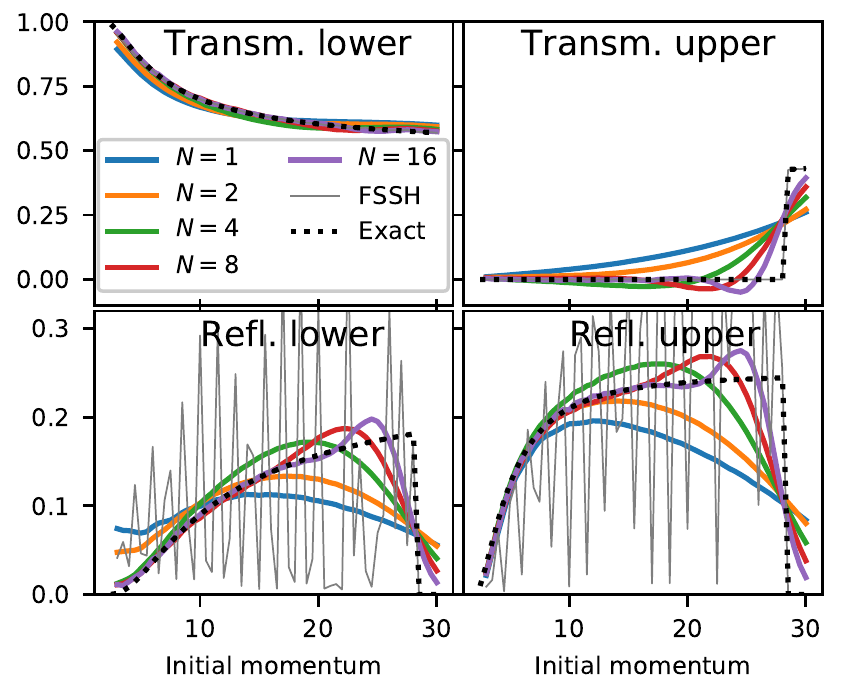}
    \caption{Transmission and reflection probabilities onto the adiabatic states in the extended coupling model (model III). While surface hopping (FSSH) suffers from erroneous oscillations, the spin path-integral method appears to be converging smoothly with increasing $N$ towards the exact scattering probabilities.} %
    \label{fig:tully3_scatt}
\end{figure}

Another well-known consequence of the overcoherence problems in surface hopping are %
erroneous oscillations \cite{subotnik2011initial} in the scattering probabilities as shown in Fig.~\ref{fig:tully3_scatt}. For the spin path-integral method, we observe that the calculated scattering probabilities converge towards the correct values with increasing $N$ (although reproducing the step as the upper channel opens appears to be difficult). Note that we did not need to add `decoherence corrections' for each trajectory (as is commonly done to fix surface hopping), but nevertheless do not observe the problems of overcoherence for the ensemble as a whole.

Finally, we note that unlike surface hopping, the results of the present method (like Ehrenfest and other mapping approaches \cite{Sun1997mapping,cotton2017adiabatic}) are not dependent on whether the adiabatic or diabatic representation is used.

\emph{Conclusions}.---In this Letter we have showed that features of quantum-classical entanglement, such as wavepacket branching and impurity measurements, can indeed be captured by an ensemble of independent and deterministic classical trajectories. This discovery opens up for a new class of mixed quantum-classical methods, as an alternative to surface hopping, coupled-trajectory or semiclassical simulations. It also extends the applicability of mapping approaches, which have been successful for predicting electronic coherences but so far have struggled to describe the nuclear dynamics of scattering problems. %
The presented method relies on positive and negative trajectory weights whose sign cancellation does not become more difficult for larger systems or longer simulation time. We therefore expect it to be applicable to complex molecular systems and condensed-phase problems.  %

Here we have limited the treatment to two-level systems, but a multi-level extension already exists for linearized spin mapping \cite{runeson2020} and the spin path-integral extension is straightforward (although there is no guarantee that $G_N$ will depend on only a scalar variable). %
Since the SW formalism can be applied to any symmetry group \cite{Tilma2016}, a similar treatment could be made also in systems with different symmetries.

Particularly interesting is the case where $\hat{\rho}$ is a thermal density matrix. Since the weights are preserved by the dynamics, we expect this to be useful for equilibrium dynamics as the quantum Boltzmann distribution will automatically be conserved. The details are left to a forthcoming paper.

\begin{acknowledgments}
The authors acknowledge support from the Swiss National Science Foundation through the NCCR MUST Network and from the Hans. H. G\"{u}nthard scholarship.
We also thank Annina Lieberherr, Joseph Lawrence, Jonathan Mannouch and Graziano Amati for fruitful discussions. %
\end{acknowledgments}

\bibliography{references,johanrefs}%

\end{document}


\title{Quantum entanglement from classical trajectories: Supplemental Material} %

\author{Johan E. Runeson}
\email{johan.runeson@phys.chem.ethz.ch}
\author{Jeremy O. Richardson}%
\email{jeremy.richardson@phys.chem.ethz.ch}
\affiliation{Laboratory of Physical Chemistry, ETH Z\"{u}rich, 8093 Z\"{u}rich, Switzerland}

\date{\today}%

\maketitle

\section{Simulation details}

\subsection{Universal function}
In the comma-separated-values file that is part of this Supplementary Material, we provide values of $4\pi\bar{s}^2 G_N(\bar{s})$ on a grid of $\bar{s}$ for $N=2$, 4, 6, 8, 12, and 16.
We obtained these results with very long Monte Carlo runs to sample $g_N(\{\bm{s}_k\})$ and histogrammed the centroid length $\bar{s}$.
To use the results in a spin path-integral simulation, we recommend interpolating it with a spline fit. %
If one samples $\bar{s}$ uniformly from $[0,\frac{\sqrt{3}}{2}]$ and the direction $\bar{\bm{s}}/\bar{s}$ uniformly from the unit sphere, then the correct weight is given by the interpolated function. %

\subsection{Model systems}
Model I was defined by the diabatic potential
\begin{subequations}\begin{align}
    &V_{11}(x) = -V_{22}(x) = A \tanh(Bx), \\
    &V_{12}(x) = V_{21}(x) = C e^{-Dx^2},
\end{align}\end{subequations}
with $A=0.01$, $B=1.6$, $C=0.005$, $D=1$. This is a slightly modified version of Tully's first model problem \cite{Tully1990hopping} in the form used by Miller and coworkers  \cite{miller2007scivr}.

Model II is Tully's dual avoided crossing model defined by the diabatic potential
\begin{subequations}\begin{align}
    &V_{11}(x) = 0, \quad V_{22}(x) = -A\,e^{-B x^2} + E_0,\\
    &V_{12}(x) = V_{21}(x) = C\, e^{-Dx^2},
\end{align}\end{subequations}
with $A=0.1$, $B=0.28$, $C=0.015$, $D=0.06$, $E_0=0.05$.

Model III was defined by the diabatic potential
\begin{subequations}\begin{align}
    &V_{11}(x) = -A, \quad V_{22}(x) = A,\\
    &V_{12}(x) = \begin{cases} B\, e^{+Cx} & x<0,\\
                     B(2-e^{-Cx}) & x>0,
                \end{cases}
\end{align}\end{subequations}
with $A=6\times 10^{-4}$, $B=0.1$, $C=0.9$. This is identical to the third of Tully's model problems \cite{Tully1990hopping} although the state labels have been swapped compared to the original paper so that 1 is always the index of the lowest state.

In each case, the mass was $m=2000$ and all quantities are reported in atomic units.

\subsection{Initial conditions}
The initial condition in all models was a product state $\hat{\rho}=\hat{\rho}_\mathrm{n}\otimes\hat{\rho}_\mathrm{e}$, where $\hat{\rho}_\mathrm{n}=|\psi\rangle\langle \psi|$ and $\hat{\rho}_\mathrm{e}=|1\rangle\langle 1|$. For the simulations shown in Figs.~2--3 of the main text, the nuclear wavefunction was a Gaussian wavepacket of the form
\begin{equation}
\psi(x)=\left(\tfrac{\gamma_0}{\pi}\right)^{1/4}\exp\left[-\frac{\gamma_0}{2}(x-x_0)^2+i p_0 (x-x_0)\right],
\end{equation}
where $x_0=-15$ is well to the left of the interaction region.
The numerically-exact wavepacket simulations were thus initiated with $|\psi\rangle\otimes|1\rangle$.

For MFT and FSSH, the initial values of $x$ and $p$ were sampled from Wigner distribution of the nuclear part,
\begin{equation}
    \rho_\text{n}(x,p)=\frac{1}{\pi}\exp\left[-\gamma_0(x-x_0)^2-\frac{1}{\gamma_0}(p-p_0)^2\right],
    \label{Wigner}
\end{equation}
together with a method-specific treatment of the electronic part $\hat{\rho}_\mathrm{e}=|1\rangle\langle 1|$. For MFT, this corresponds to all trajectories starting with a spin at the north pole of a sphere of radius $\thalf$. %

\begin{figure}
    \centering
    \includegraphics{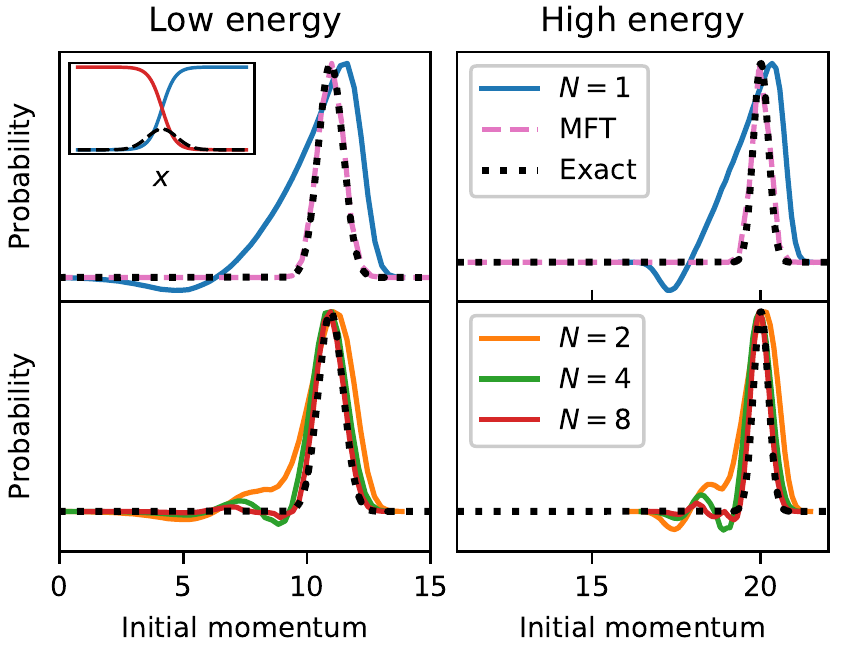}
    \caption{Initial momentum distribution in model I.}
    \label{fig:init_p}
\end{figure}

Note that the spin path-integral dynamics are initialized from all parts of the spin sphere, in contrast with Ehrenfest and FSSH\@. Therefore, in order for the total energy distribution to be consistent with the wavepacket calculation, the initial conditions need to take account of the sampled value of $V(\bs)$.
It is necessary to do this because the same set of spin path-integral trajectories can in principle be used to study the alternative scattering problem with the same total energy distribution initialized on the upper state.
The simplest algorithm for obtaining the appropriate energy distribution is to sample a momentum $p'$ from the distribution Eq.~\eqref{Wigner} to obtain $E=\frac{(p')^2}{2m}+V_{11}$.
Then the corrected momentum $p$ is defined by the solution of $E=\frac{p^2}{2m}+V(\bs)$, and this is used to initialize the trajectory dynamics.
Although this modification appears to change the initial momentum distribution, it can be seen that in the $N\rightarrow\infty$ limit the resulting distribution is formally correct (see Figure~\ref{fig:init_p}) due to the fact that $V(\bs)\rightarrow V_{11}$ on account of the peaking of the distribution of $\bs$.
Simple extensions of this algorithm can be constructed to treat multidimensional systems by sampling a distribution of $E$ and the direction of the momentum vector.

In all spin path-integral dynamics calculations, we used the single-variable universal function $G_N(\bar{s})$ rather than the more complicated $g_N(\{\bm{s}_k\})$.  (As shown in the main text, in principle they both give the same result although the former gives better numerical convergence). The initial electronic state was sampled together with this function, meaning that $\bs$ was sampled by Monte Carlo from the absolute value of the distribution
\begin{align}\label{eq:bs_distr}
    f(\bs) \, d^3\bs &= G_N(\bar{s}) \rho_\text{e}(\bs) \, d^3\bs \\
    &= 4\pi\bar{s}^2G_N(\bar{s})\left(\tfrac{1}{2}+\bar{s}_z\right) d\bar{s}\, \frac{\sin\theta\, d\theta \, d\varphi}{4\pi}
\end{align}
and weighted by its sign. %

In Fig.~2 of the main text (Model I) we considered two different initial wavepackets: one with lower energy $\frac{p_0^2}{2m}=0.03$ with $\gamma_0=0.5$ and one with higher energy $\frac{p_0^2}{2m}=0.1$ with $\gamma_0=0.1$.
In Fig.~3 of the main text (Model III) we used $p_0=10$ and $\gamma_0=0.5$.

The scattering calculations in Figs.~4 (Model III) and~\ref{fig:tully2} (Model II) were initiated with $\rho_\text{n}(x,p)={\delta(x-x_0)\delta(p-p_0)}$ with a range of $p_0$ values (because we are computing only scattering probabilities, the initial $x$ distribution is unimportant in this case and chosen merely for simplicity).
The reason we chose to include results from calculations with a well defined initial momentum is that a distribution of initial values of $p$ is known to fortuitously smear out the oscillations observed in other methods like FSSH, without solving the underlying problem of overcoherence \cite{subotnik2011initial}. As before, the initial momentum in the spin path-integral simulations was corrected to reproduce the correct initial energy.

\subsection{Dynamics and observables}
Although it is possible to evolve the $\bs$ dynamics directly, it is simpler (and gives more stable algorithms) to transform to Cartesian %
coordinates $X_1,P_1,X_2,P_2$ via
\begin{subequations} \label{eq:ntoXP}
\begin{align}
    2\bar{s}_x &= X_1 X_2 + P_1 P_2 \\
    2\bar{s}_y &= X_1 P_2 - X_2 P_1   \\
    2\bar{s}_z &= \tfrac{1}{2}(X_1^2 + P_1^2 - X_2^2 -P_2^2).
\end{align}
\end{subequations}
As in Ref.~\onlinecite{runeson2019}, one can show that the equations of motion $\dot{\bs}=\bs\times\bm{H}$ and Eq.~(4) in the main text %
are equivalent to to Hamilton's equations of motion with the Hamiltonian
\begin{multline}
        H = \frac{p^2}{2m} + U(x) + \sum_{n=1}^2 V_{n}(x)\frac{1}{2}(X_n^2+P_n^2-\gamma)\\
        +\Delta(x)(X_1 X_2 +P_1 P_2)
\end{multline}
with conjugate variables $(x,p)$, $(X_1,P_1)$ and $(X_2,P_2)$. Here, we assumed that the diabatic potential matrix, $\hat{V}(x)$, is chosen to be real. The term $\gamma$, which for historical reasons is called the \emph{zero-point energy parameter} \cite{Mueller1999pyrazine}, is given by
\begin{equation}
    \gamma = 2\bar{s}-1.
\end{equation}
For $N=1$, where $\bar{s}=\tfrac{\sqrt{3}}{2}$ is fixed, this reduces to the value $\sqrt{3}-1$ that is familiar from Refs.~\onlinecite{Cotton2013b,runeson2019}. For ${N>1}$, $\gamma$ is different for each trajectory and always smaller than the $N=1$ value.

All spin path-integral simulations were carried out in the diabatic representation. For model III, results are reported in the adiabatic representation. The diabatic to adiabatic transformation is done separately for each trajectory. The impurity was calculated by first averaging each element of the reduced density matrix in the adiabatic basis and then post-processing the results to evaluate  $S_L=2(\rho_{11} \rho_{22}-|\rho_{12}|^2)$.%

FSSH trajectories were propagated on the adiabatic surfaces using frustrated hops and without including any decoherence corrections.
Surface-hopping scattering probabilities were computed from the final distribution of active states, while the impurity measurement shown in Fig.~3 was computed from the reduced density matrix in the adiabatic basis, as for the other methods. %

Wavepackets were propagated using the split-operator approach \cite{Tannor}
and numerically-exact scattering probabilities were obtained using the log-derivative method \cite{Mano1986logderivative,Thornley1994logderivative,Alexander2001diabatic}.

\section{Additional Results}
\subsection{Results for model II}
\begin{figure}
    \centering
    \includegraphics{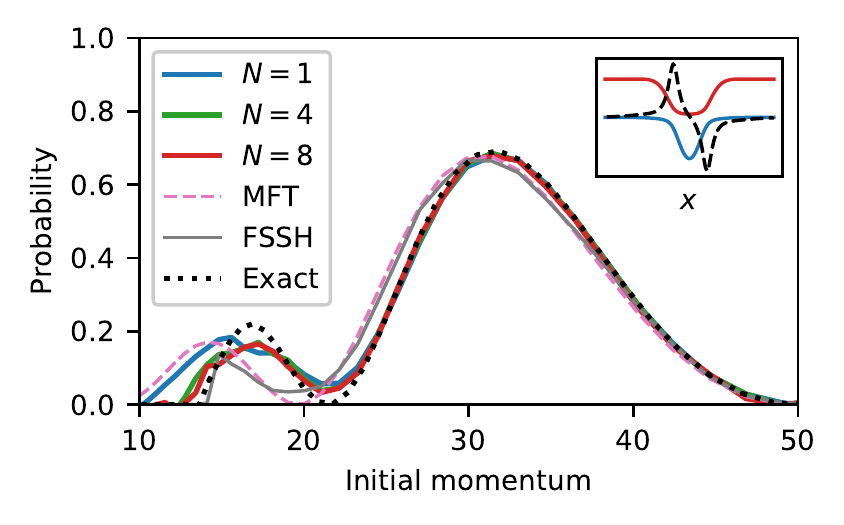}
    \caption{Probability of transmission on the upper state in the dual avoided crossing model (model II). Inset shows adiabatic surfaces (solid lines) and nonadiabatic coupling (dashed line, divided by 12). Our surface hopping (FSSH) results are in agreement with the equivalent calculations of Ref.~\cite{subotnik2011AFSSH}.}
    \label{fig:tully2}
\end{figure}

Figure~\ref{fig:tully2} shows the transmission probability onto the upper state for the dual avoided crossing (model II). As in the other models, the initial state was on the lower surface.

The results show that the linearized spin-mapping method ($N=1$) is already very accurate for this problem except in the low-energy regime and that the remaining error is significantly decreased when using the path-integral spin-mapping for $N>1$.
The conclusions from this model are thus in line with those discussed in the main text.

\begin{figure}
    \centering
    \includegraphics{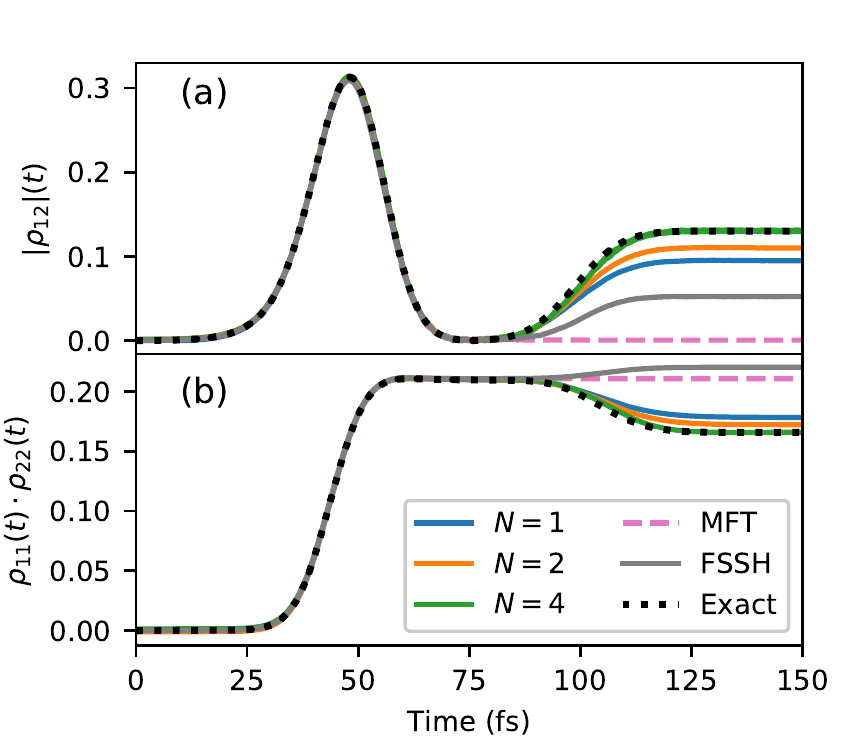}
    \caption{Decoherence indicators in the extended coupling model. The off-diagonal element of the reduced density matrix (panel a) and the product of populations (panel b) are both correctly reproduced by the spin path-integral method, whereas Ehrenfest (MFT) and surface hopping (FSSH) both fail to describe the second crossing.}
    \label{fig:tully3_terms}
\end{figure}

\subsection{Analysis of the impurity in Model III}
Figure~\ref{fig:tully3_terms} shows separately the quantities $|\rho_{12}|(t)=\sqrt{(\mathrm{Re}\,\rho_{12}(t))^2+(\mathrm{Im}\,\rho_{12}(t))^2}$ and $\rho_{11}(t)\cdot\rho_{22}(t)$ that are used in computing the impurity in the main text. (All quantities refer to the reduced density matrix in the adiabatic representation.) The off-diagonal element is a direct measure of coherence, and the second crossing gives rise to a \emph{recoherence} that is correctly described by spin path-integral method, but not by Ehrenfest or surface hopping.
These decoherence indicators are related to that used in Ref.~\onlinecite{Min2015nonadiabatic} to test the quantum-classical coupled-trajectory solutions of the exact factorization formalism.

\subsection{Convergence details}

\begin{figure}
    \centering
    \includegraphics{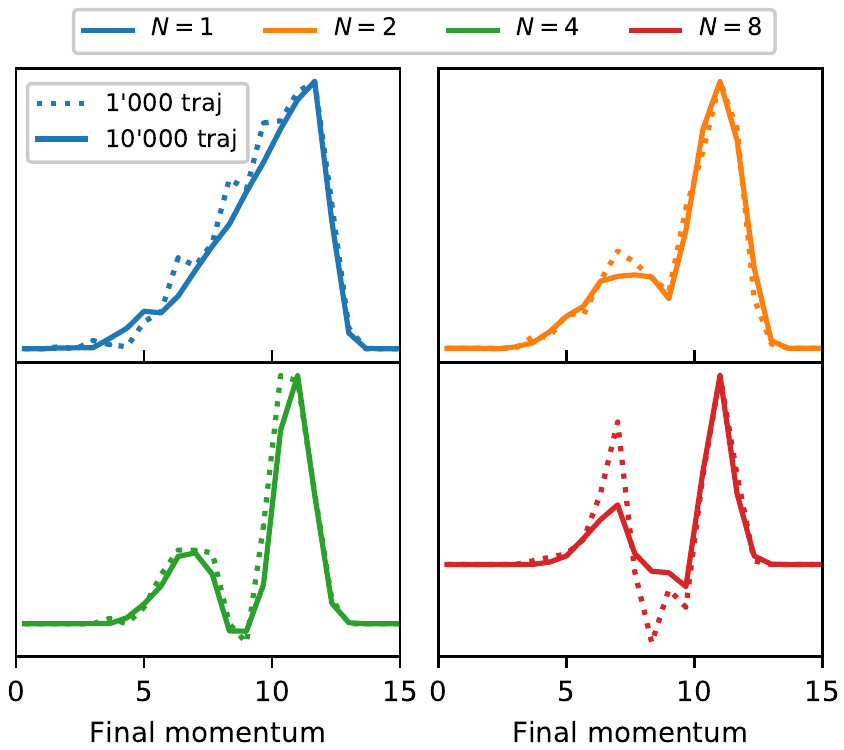}
    \caption{Final momentum distribution in Model I for an initial wavepacket centred around an energy of ${p_0^2}/{2m}=0.03$. Higher $N$ generally requires more trajectories for convergence although the two-peak structure can already be observed with only 1000 trajectories. }
    \label{fig:conv}
\end{figure}

\begin{figure}
    \centering
    \includegraphics{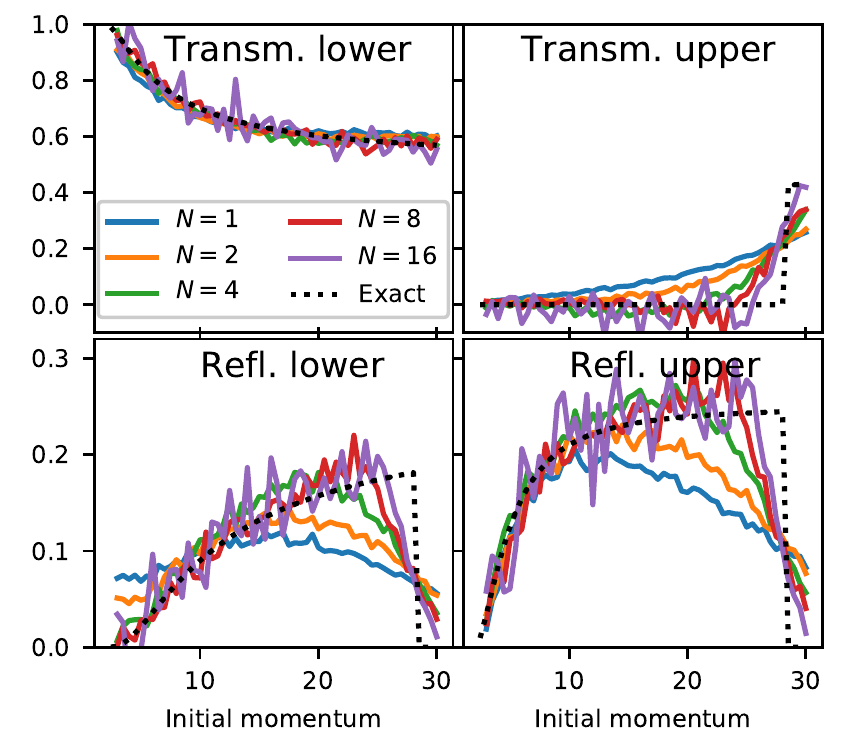}
    \caption{Scattering probabilities in Model III (extended coupling model) with $10^4$ trajectories for each initial momentum.}
    \label{fig:tully3_1e4}
\end{figure}

The number of trajectories used in Model I was $10^6$ in order to provide fully-converged high-resolution histograms, but the overall trend can be observed already with $10^3$ trajectories (i.e.\ just tens of trajectories per histogram bin), see Fig.~\ref{fig:conv}. Converged results for model II were computed from $10^5$ trajectories. All spin-dynamics calculations for model III used $10^6$ trajectories, although the important trends can already be clearly seen with just $10^4$, see Fig.~\ref{fig:tully3_1e4}.

The convergence rate is mostly determined by the average sign of $f(\bs)$ [Eq.~\eqref{eq:bs_distr}], which is reported in Table~\ref{tab:sign}.  It is clear that it is far more efficient to use the single-variable universal function $G_N(\bar{s})$ instead of the path-integral weight $g_N(\{\bm{s}_k\})$, especially as $N$ increases.  These results are universal for any two-level quantum-classical system given a pure initial state in the quantum system, and as stated in the main text, independent of both the trajectory length and the number of environmental degrees of freedom. The values imply that typically a simulation with $N=16$ will require only one order of magnitude more trajectories than $N=1$ for the same level of convergence.

\begin{figure}
    \centering
    \includegraphics{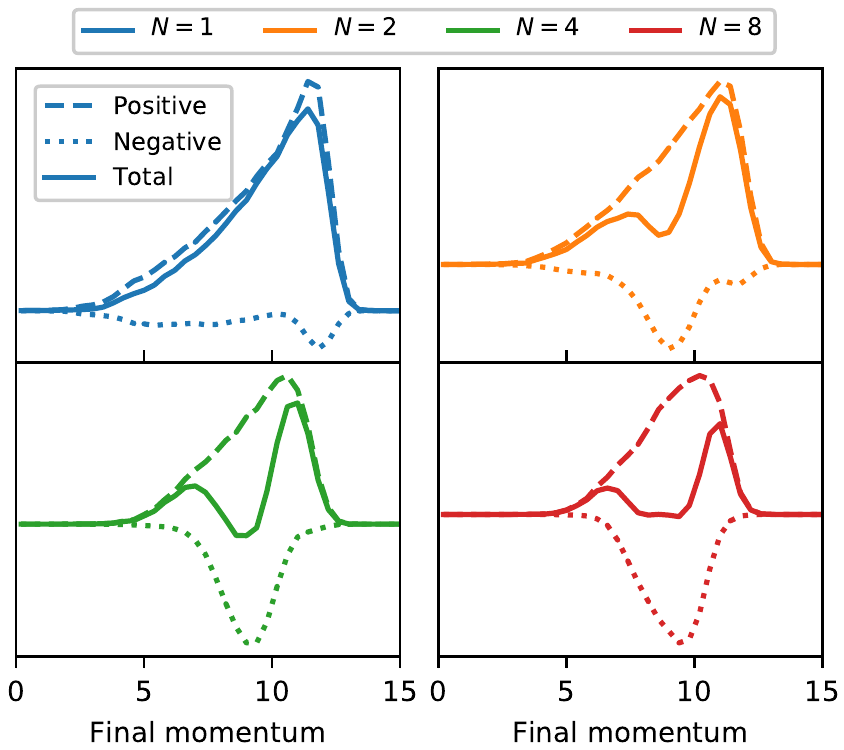}
    \caption{Final momentum distribution in Model I for an initial energy $\frac{p_0^2}{2m}=0.03$. Cancellation of positively weighted (dashed line) and negatively weighted (dotted line) samples is what yields the two-peak structure (solid line).}
    \label{fig:signed}
\end{figure}

Fig.~\ref{fig:signed} shows the contributions of positively and negatively weighted trajectories separately, and it is clear that the two-peak structure emerges from cancellation of trajectories with intermediate momenta.

\begin{table}
    \centering
    \begin{ruledtabular}
    \begin{tabular}{cd{-1}d{-1}d{-1}d{-1}d{-1}}
        $N$ & \head{1} & \head{2} & \head{4} & \head{8} & \head{16} \\ \hline
        $\langle \mathrm{sgn}\rangle_{G_N}$ & 0.87 & 0.56 & 0.32 & 0.19 & 0.11  \\
        $\langle \mathrm{sgn}\rangle_{g_N}$ & 0.87 & 0.56 & 0.20 & \sim 0.017 & < 0.001
    \end{tabular}
    \end{ruledtabular}
    \caption{Top row: average sign when sampling $f(\bs)$ in Eq.~\eqref{eq:bs_distr} for various $N$. Bottom row: the corresponding average sign if one instead integrated over $N$ spin variables weighted by $g_N(\{\bm{s}_k\})$. In our simulations we use $G_N(\bar{s})$, corresponding to the top row. These values are model-independent and valid for any pure initial state.}
    \label{tab:sign}
\end{table}

\clearpage

\bibliography{references,johanrefs}